\documentclass{article}
\usepackage{spconf,amsmath,graphicx}
\usepackage{bbm}
\usepackage{amssymb}
\usepackage{multirow}
\usepackage{makecell}
\usepackage{url}

\title{Radar Emitter Classification with \\ Attribute-specific Recurrent Neural Networks}

\name{Paolo Notaro$^{1,2}$ \qquad Magdalini Paschali$^1$ \qquad Carsten Hopke$^2$ \qquad David Wittmann$^2$ \qquad Nassir Navab$^{1,3}$}
\address{
        $^1$ Computer Aided Medical Procedures, Technical University of Munich, Germany\\
        $^2$ Airbus Defence and Space, Manching, Germany\\
        $^3$ Computer Aided Medical Procedures, Johns Hopkins University, Baltimore, USA}

\begin{document}
%
\maketitle
\begin{abstract}
Radar pulse streams exhibit increasingly complex temporal patterns and can no longer rely on a purely value-based analysis of the pulse attributes for the purpose of emitter classification. In this paper, we employ Recurrent Neural Networks (RNNs) to efficiently model and exploit the temporal dependencies present inside pulse streams. With the purpose of enhancing the network prediction capability, we introduce two novel techniques: a per-sequence normalization, able to mine the useful temporal patterns; and attribute-specific RNN processing, capable of processing the extracted information effectively. The new techniques are evaluated with an ablation study and the proposed solution is compared to previous Deep Learning (DL) approaches. Finally, a comparative study on the robustness of the same approaches is conducted and its results are presented.
\end{abstract}

\section{Introduction}
\label{sec:intro}
Emitter classification of radar pulse sequences is a critical task in Electronic Warfare (EW) disciplines such as Electronic Support Measures (ESM), where the correct identification of a target is crucial to determine adequate countermeasures for protection of sensible units and other defence purposes~\cite{adamy, spezio}. The increasing complexity of the electromagnetic environment, due to more sophisticated radar characteristics and higher emitter density, has rendered classification of pulse streams an increasingly difficult task~\cite{spezio}.

Traditional approaches to tackle this problem are based on categorization of pulses through statistical measures of different pulse attributes~\cite{wiley}. Commonly exploited pulse features are, in this sense, the Pulse Width (PW) and the Radio Frequency (RF). During a process known as deinterleaving, incoming pulses are clustered by emitter~\cite{adamy, deinterleaving} so that time-variant parameters, like the pulse repetition interval (PRI), are computed and can be further utilized for classification~\cite{pri-classification}. Other intrapulse features are occasionally used, although frequently discarded in real-time operations to avoid storage overloading. More recent is the application of Machine Learning and especially Deep Learning (DL) to radar pulse stream classification, for which proposed solutions include Support Vector Machines (SVMs)~\cite{svm}, Multilayer Perceptrons (MLPs)~\cite{petrov, shieh} and Convolutional Neural Networks (CNNs)~\cite{cnn-hong, cnn-cain, cnn-sun}. 

However, these approaches have several shortcomings. Firstly, a sufficient number of pulses needs to be acquired before taking a prediction step, having repercussions on their real-time applicability. Secondly, temporal patterns and dependencies inside pulse streams are not modelled efficiently, because the pulse order inside the sequence is either not taken into account or, in some cases, the entire series of values is summarized to the average or the domain interval of the attributes~\cite{petrov}. Towards the overcoming of the aforementioned shortcomings, a recent work~\cite{radar-rnn} has introduced Recurrent Neural Networks (RNNs) in the ESM domain as a method to efficiently process pulse streams, due to their proven effectiveness in several sequence processing problems, such as neural machine translation, time series forecasting and classification~\cite{nmt, forecasting, tsc-survey}.

In this paper, we propose to utilize attribute-specific RNNs in combination with a novel normalization scheme, for the challenging task of emitter classification. Towards this end our contribution is two-fold: 1) we introduce a new type of normalization, here called per-sequence normalization, and we apply it in parallel to the more commonly used min-max normalization, concatenating the output of the two transformations along the feature axis to obtain $2 * M$ channels, where $M$ is the number of attributes extracted from each pulse. 2) we leverage attribute-specific Long Short Term Memory (LSTM) layers~\cite{lstm} for each feature after the normalization process to compute an intermediate representation useful for classification purposes (see Fig.~\ref{fig:architecture}).

\section{Methodology}
\label{sec:methodology}


Our method utilizes attributes that are extracted from Pulse Descriptor Words (PDWs) to construct sequences of pulses. These sequences are of the form $\boldsymbol{S} = [\boldsymbol{s_1}, \boldsymbol{s_2}, ..., \boldsymbol{s_t}, ..., \boldsymbol{s_T}]$, $T$ being the length of the sequence, where the individual pulse is represented as a tuple $\boldsymbol{s_t} = (s^1_t, s^2_t, ..., s^j_t, ..., s^M_t) \in \mathbb{R}^M$, $M$ being the number of pulse attributes extracted from the PDW. Moreover, we define $S^j = [s^j_1, s^j_2, ..., s^j_t, ..., s^j_T]$ as the $j$th-attribute sequence of $\boldsymbol{S}$. Finally, each sequence $\boldsymbol{S}$ is associated with a class $y$, $1\leq y \leq C$, to form a dataset $\mathcal{D} =\{(\boldsymbol{S}_1, y_1), (\boldsymbol{S}_2, y_2), ..., (\boldsymbol{S}_i, y_i), ..., (\boldsymbol{S}_N, y_N)\}$ of $N$ samples divided in $C$ classes. For a classifier $f_w(\bullet)$ with output prediction $\hat{y}=f_w(\boldsymbol{S})$, the goal is to maximize the classification accuracy, so that $\hat{y}_i = y_i$ for as many $\boldsymbol{S}_i,y_i \in \mathcal{D}$ as possible. The classifier $f_w$ is parametrized by its weights $w$, which are subject to optimization and trained via first-order methods based on gradient descent.
Finally, the model employs LSTM~\cite{lstm} cells, which have shown to be able to learn long-term dependencies (by means of the internal cell state) while handling the vanishing and exploding gradient problems typically encountered with standard RNN cells~\cite{rnn-gradient}.

\subsection{Normalization scheme} The model $f_w$ incorporates a normalization scheme that maps the original sequence $\boldsymbol{S}$ to $\overline{\boldsymbol{S}}$. Differently from~\cite{radar-rnn}, the input is not digitized but normalized according to two different techniques. The first one is min-max normalization, for which the sequence values are linearly mapped into the range $[-1, 1]$ according to:

\begin{equation*}
\overline{S}^j = 2\cdot\frac{S^j - MIN_j(\mathcal{D})}{MAX_j(\mathcal{D}) - MIN_j(\mathcal{D})} - 1,\;\forall j
\end{equation*}

The attribute domains $[MIN_j(\mathcal{D}), MAX_j(\mathcal{D})]$ are estimated from the whole training data distribution and the normalization is applied attribute-wise to all the sequences of the dataset.
The second transformation applied is a per-sequence normalization, i.e.\ sequence attributes are normalized based on the values inside their respective attribute sequence only. This differs from min-max normalization, where the entire dataset distribution of values plays a role in defining the normalization. As $S^j$ defines the values of attribute $j$ in sequence $\boldsymbol{S}$, we have:

\begin{equation*}
\overline{S}^j = 2\cdot\frac{S^j - \min_t(S^j)}{\max_t(S^j) - \min_t(S^j)} - 1,
\end{equation*}

This normalization is again applied for $\;1\leq j \leq M$ in the sequence and for all $\;1\leq i \leq N$ in the dataset independently, mapping the observed attribute sequence domain to the range $[-1, 1]$. Doing so enables the network to isolate temporal patterns inside the sequence with more precision, due to the restricted domain of values taken by attribute along the time axis only. The two normalizations are performed in parallel and the outputs are concatenated along the feature axis to obtain a $T \times 2M$ input. Compared to the discretization proposed in~\cite{radar-rnn}, this normalization scheme allows for reduced model complexity and increased inference speed, by eliminating the need for embeddings and reduction of the input dimension of the remaining network.

\begin{figure}
    \centering
    \resizebox{\columnwidth}{!}{\includegraphics{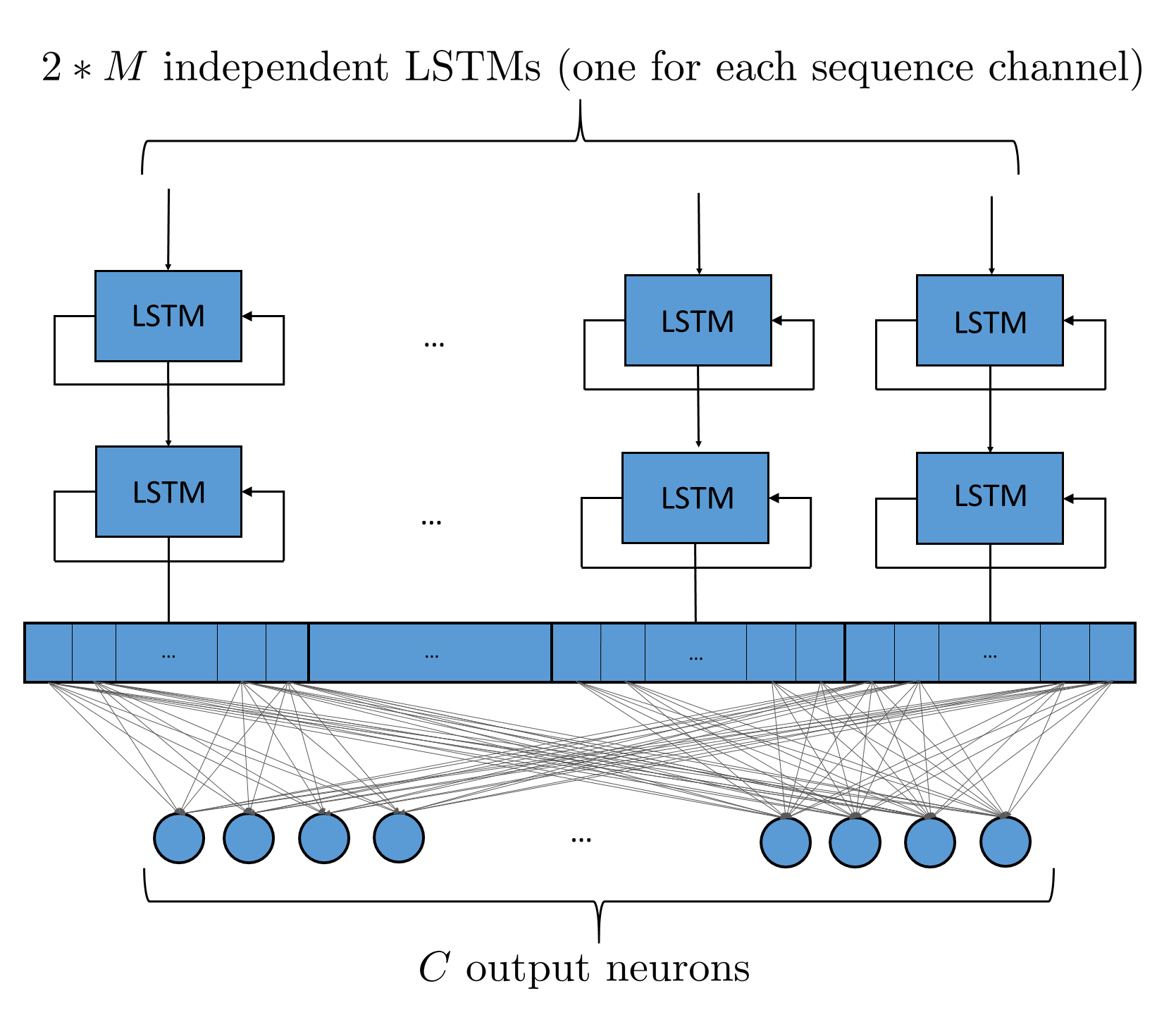}}
    \caption{Proposed network architecture.}
    \label{fig:architecture}
\end{figure}

\subsection{Attribute-specific LSTMs} Other differences with respect to~\cite{radar-rnn} are the choice and the architectural layout of the RNN layers. In our approach, a dedicated RNN network is assigned to each sequence channel, so that temporal dependencies can be extracted, while the values of different attribute sequences can be processed separately (Fig.~\ref{fig:architecture}). This is motivated by the fact that joint-attribute patterns are oftentimes spurious since attribute values changes occur independently. These RNN blocks are constituted of $L$ stacked LSTM~\cite{lstm} layers, each block producing an output feature map of size $T \times h_{LSTM}$. After the RNN layers have processed the normalized input sequences, the outputs of the $2 * M$ LSTM layers are concatenated along the hidden size dimension to obtain a $T \times 2 * h_{LSTM} * M$ output sensor for a $T$-long input sequence of $M$ pulse attributes. In our experiments we set $L=2$ and $h_{LSTM}=64$.

\subsection{Model training} The architecture is completed with a Fully Connected (FC) layer, mapping the hidden features from the LSTMs to the prediction class scores. The softmax function is then applied on these prediction scores to obtain the final class probabilities $\boldsymbol{\hat{y}_i} \in \mathbb{R}^C$, which represent the input of the loss function. The loss function employed for training is the weighted Cross Entropy loss~\cite{weighted-cross-entropy}, defined as

\begin{equation*}
   L = -\frac{1}{W}\sum_{i=1}^N w_{y_i} \log(\hat{y}_{i,y_i}),\;W = \sum_{i=1}^N\frac{1}{w_{y_n}}
\end{equation*}

where weights $w_c$, $1\leq c\leq C$ are estimated on the training set through median frequency balancing~\cite{mfb}. Finally, dropout~\cite{dropout} (with $p=0.5$) is used between the stacked LSTM layers and before the final FC layer to prevent overfitting and improve generalization.

\section{Experimental Setup}
\label{sec:experimental_setup}

To test the described supervised approach, a dataset of labeled pulse sequences was produced. It was obtained by means of a radar simulation software, in which different radar characteristics were defined and where, resembling a real-world scenario, signals were programmed to be detected by the receiving antenna of an external agent. Afterwards detected pulses are clustered by emitter and sorted by time, similar to a real-operating threat detection system, to produce the final pulse sequences. The dataset consists of 60910 training samples and 17382 test samples, divided into 17 classes of emitters, taken from different operating environments (aircraft, marine or ground-based). To depict a more realistic case, classes are not equally represented inside the dataset, making the classification task more challenging. Moreover, the pulse sequences present in the dataset have variable length, ranging from few pulses (7-8) to an upper limit of 512. In our experiments, the set of pulse attributes consists of PRI, PW and RF.

The class imbalance was taken into consideration in terms of evaluation metrics as well. To emphasize correct classification for all the classes, regardless of their relative frequency, macro-averaged classification accuracy ($M$) was measured during all the experiments. Macro-averaged metrics are obtained by computing the metric independently for each class and then taking the average, i.e.\ $M = \sum_{c=0}^{C-1}{ACC_c}/C$, where $ACC_c=\sum_{i=1}^N\mathbbm{1}(y^{pred}_i=y_i=c)/N_c$. Accuracies have been measured on the test set, consisting of unseen examples, in order to test the generalization power of the models.

This evaluation metric was used to perform an ablation study of the different techniques introduced in the paper. First, a model was tested without any input normalization. Then the same model was tested with min-max input normalization only. Finally, the same evaluation was performed on a model with the proposed normalization scheme. Results are then compared. According to the same principle, the same model was tested with and without attribute-specific LSTM layers to showcase their effectiveness.

Moreover, a baseline comparison of different DL-based emitter classification approaches was carried out on our dataset. Other tested models include:
\begin{itemize}
\item The already mentioned Liu et al.~\cite{radar-rnn}, who apply a RNN architecture based on Gated Recurrent Units (GRUs)~\cite{gru} with discretization of the two input features PRI and PW, without employing RF;
\item The work of Petrov et al.~\cite{petrov}, who employ a MLP on statistics computed from attribute sequences, more specifically minimum and maximum observed values for PRI, PW, and RF;
\item A ResNet18~\cite{resnet} model, which is a state-of-the-art CNN architecture. Even though it was initially designed for image classification on 2D inputs, it has been shown to work effectively on time series classification as well~\cite{tsc-survey, tsc-resnet}.
\end{itemize}
To ensure a more objective baseline for comparison, the same set of attributes, consisting of PRI, PW and RF, was utilized. Therefore, methods who explicitly mention to use only a subset of these three pulse features were tested both in the configuration described in the original paper and in the configuration of this paper. Finally, in case some papers proposed more than one normalization, all the different proposals were included in our baselines.

\section{Results and Discussion}
\label{sec:results_discussion}

\begin{table}
    \resizebox{\columnwidth}{!}{
        \begin{tabular}{|c|c|c|c|} 
        \hline
        \multicolumn{2}{|c|}{\textbf{Normalizations}} & \multirow{2}{*}{\makecell{{\small \textbf{Attribute-specific}}\\ {\small \textbf{LSTM}}}} & \multirow{2}{*}{\textbf{$M$-accuracy}}\\ \cline{1-2}
        
        min-max                     & per-seq.                  &           &                   \\ \hline
        \multirow{2}{*}{$\circ$}    & \multirow{2}{*}{$\circ$}  & $\circ$   & 0.1445     \\ \cline{3-4}
                                    &                           & $\bullet$ & \textbf{0.1449}   \\ \hline
        \multirow{2}{*}{$\bullet$}  & \multirow{2}{*}{$\circ$}  & $\circ$   & 0.5894            \\ \cline{3-4}
                                    &                           & $\bullet$ & \textbf{0.6006}   \\ \hline
        \multirow{2}{*}{$\bullet$}  & \multirow{2}{*}{$\bullet$}& $\circ$   & 0.5030            \\ \cline{3-4}
                                    &                           & $\bullet$ & \textbf{0.6498}   \\ \hline
        \end{tabular}
    }
    \caption{Macro-averaged test accuracy for emitter classification networks with and without attribute-specific LSTMs along with different normalization schemes.}
    \label{tab:accuracies-ablative-sfp}
\end{table}

Table~\ref{tab:accuracies-ablative-sfp} summarizes the results of the ablation study regarding the prediction accuracy, when applying attribute-specific LSTMs compared to the standard joint RNN processing case. In all the cases the introduction of attribute-specific LSTMs outperforms the joint RNNs by 2-14\%. Additionally, Table~\ref{tab:accuracies-ablative-sfp} clearly highlights the improvement in accuracy of attribute-specific LSTMs brought upon by the proposed normalization scheme, rendering the combination of those two components the most favorable option for the radar emitter classification.

\begin{table}
    \centering
    \resizebox{\columnwidth}{!}{
        \begin{tabular}{|c|c|c|c|c|}
            \hline
            \textbf{Method}                     & \textbf{Normalizations}   & \textbf{$M$-accuracy}   \\ \hline
            Liu et al.~\cite{radar-rnn}         & discretization            & 0.4563                    \\ \hline
            Liu et al.~\cite{radar-rnn} + RF    & discretization            & 0.5477                    \\ \hline
            Petrov et al.~\cite{petrov}         & min-max                   & 0.6287                    \\ \hline
            Petrov et al.~\cite{petrov}         & standardization           & 0.5218                    \\ \hline
            Resnet18~\cite{resnet}              & min-max                   & 0.5060                    \\ \hline
            \textbf{Proposed}                   & \textbf{min-max+per-seq.} & \textbf{0.6498}           \\ \hline
        \end{tabular}
    }
    \caption{macro-averaged test accuracy for emitter classification networks of different approaches.}
    \label{tab:accuracies-comparison}
\end{table}


\begin{figure}
    \centering
    \resizebox{\columnwidth}{!}{\includegraphics{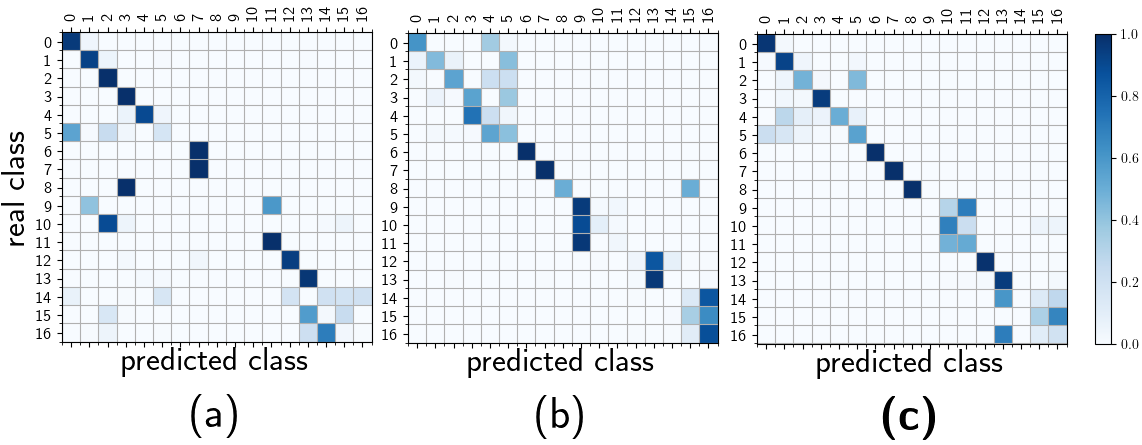}}
    \caption{Confusion matrix of the test set accuracy for different approaches: (a) Liu et al.~\cite{radar-rnn}, (b) ResNet18~\cite{resnet}, \textbf{(c) Proposed}.}
    \label{fig:cf_comparison}
\end{figure}

In Table~\ref{tab:accuracies-comparison} the accuracy of our method is compared with a variety of approaches which have been previously deployed for emitter classification. The superiority of the proposed method is clear with an improvement across the board ranging between 2\%-19\%. Specifically, comparing our method with Liu et al.~\cite{radar-rnn} the improvement brought upon our method is 19\% if we utilize the method described explicitly in~\cite{radar-rnn} and 10\% when we also incorporate the RF information for fairness. The increase in performance can be attributed to the fact that the proposed normalization is more suitable for this domain compared to discretization. 

Furthermore, we achieved a 2\%-12\% improvement in comparison to Petrov et al.~\cite{petrov}. Utilizing temporal information and per-sequence normalization, which isolates temporal patterns inside the sequences efficiently, provides a tailored approach for the emitter classification. Finally, a state-of-the-art ResNet18~\cite{resnet} was also outperformed by our method by 14\%, highlighting the importance of the temporal information stored by the LSTMs.

In Figure~\ref{fig:cf_comparison}, we compare the confusion matrices of Liu et al.~\cite{radar-rnn} and ResNet18~\cite{resnet} with the proposed method, in order to showcase the performance achieved per class and the main sources of ambiguity that increased the difficulty of the task at hand. Neighboring classes usually represent similar type of emitters, such as air-based or ground-based. Thus we can clearly see that all methods are prone to confusing classes originating from similar emitter types. However, the proposed method achieves the lowest amount of misclassifications and as can be seen the majority of the predictions lie in the diagonal overlapping with the ground truth classes.


\begin{figure}
    \centering
    \resizebox{\columnwidth}{!}{\includegraphics{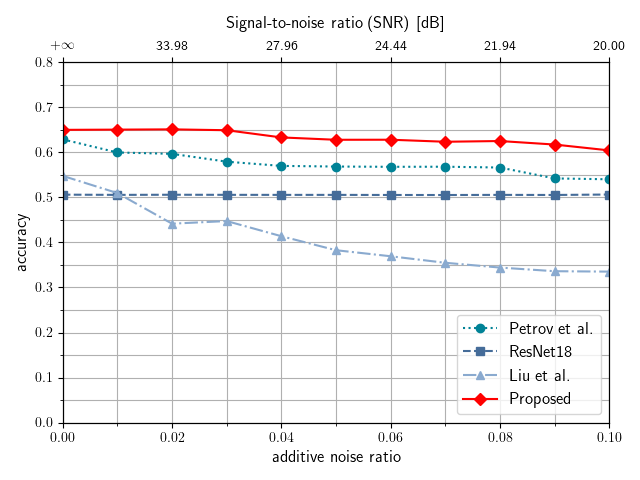}}
    \caption{Robustness to additive noise vs. accuracy for different approaches.}
    \label{fig:robustness-noise}
\end{figure}

Finally, we performed a robustness evaluation experiment by adding Gaussian noise on the radar signals in order to showcase the resistance of our method to noise perturbations. In our experimental setup, additive Gaussian noise was applied in increasing quantities to the pulse sequences reaching up to 10\% of the signal magnitude. This value corresponds to a Signal-to-Noise Ratio (SNR) of 20 dB, a threshold in the proximity of the minimum SNR requirement for most EW systems~\cite{adamy}. Fig.~\ref{fig:robustness-noise} shows that all the methods, except for Liu et al.~\cite{radar-rnn} maintained a relatively stable performance in presence of noise. However, the proposed approach still outperformed the other baselines increasingly as the percentage of noise got higher, ranging from 2\% increase for no noise to 6\% for a SNR of 20 dB in comparison to Petrov et al.~\cite{petrov}. ResNet18~\cite{resnet} shows high robustness to noise, however the proposed method combines not only resilience to noise but also an improved accuracy by 15\%.  

\section{Conclusion}
\label{sec:conclusion}

In this paper we introduced a novel method for the complex task of radar emitter classification. Our approach comprises an application-specific normalization scheme to address the large variability of values within signal attributes and feature-specific RNNs, which not only incorporate temporal dependencies in the method but also improve the processing of individual features. 

Through thorough ablation testing of the individual components of the proposed technique and comparison with previous state-of-the-art methods, we showcased the superiority of our method in terms of accuracy. Furthermore, our robustness evaluation with additive Gaussian noise proved the ability of our approach to be more stable in the presence of noise compared to other baselines. 

Future work may include the application of the proposed method across domains on other pulsed signals, such as LIDAR signal classification in autonomous driving. Other future applications expand to tasks of the medical domain, such as pulse irregularity detection in ECGs or MRIs and EEG signal classification.

\vfill\pagebreak

\bibliographystyle{IEEEbib}
\bibliography{refs}

\end{document}